\documentclass[]{llncs}

\usepackage[paperVersion]{marcpaper}
\usepackage{graphicx}
\usepackage{etex}
\usepackage{lscape}

\newcommand{\ttwo}{{\tool{T2}}\xspace}

\begin{document}
\title{\tool{T2}: Temporal Property Verification}
\author{
Marc Brockschmidt\inst{1} 
\and Byron Cook\inst{2}        
\and Samin Ishtiaq\inst{1}     
\and Heidy Khlaaf\inst{2}      
\and Nir Piterman\inst{3}
}

\institute{
  \hspace{-1ex}Microsoft Research Cambridge,
  \stepcounter{@inst}$^2$University College London,
  \stepcounter{@inst}$^3$University of Leicester
}

\makeatletter
\renewcommand\paragraph{\@startsection{paragraph}{4}{\z@}%
                       {-2\p@ \@plus -1.33\p@ \@minus -1.33\p@}%
                       {-0.25em \@plus -0.11em \@minus -0.05em}%
                       {\normalfont\normalsize\itshape}}
\makeatother

\maketitle

\vspace{-4ex}
\begin{abstract}
We present the open-source tool \tool{T2}, the first public release from the
\tool{TERMINATOR} project~\cite{Cook06a}.
\tool{T2} has been extended over the past decade to support automatic
temporal-logic proving techniques and to handle a general class of
user-provided liveness and safety properties.
Input can be provided in a native format and in \tool{C}, via the support of
the \tool{LLVM} compiler framework.
We briefly discuss \tool{T2}'s architecture, its underlying techniques,
and conclude with an experimental illustration of its competitiveness and
directions for future extensions.
\end{abstract}
\vspace{-2ex}


\vspace{-2ex}
\section{Introduction}
\label{sect:Introduction}

We present \tool{T2} (\tool{TERMINATOR} 2), an
open-source framework that implements, combines, and extends
techniques developed over the past decade aimed towards the verification of temporal
properties of programs.
\tool{T2} operates on an input format that can be automatically extracted from
the \tool{LLVM} compiler framework's intermediate representation,
allowing \tool{T2} to analyze programs in a wide range of programming languages
({\em e.g.} \tool{C}, \tool{C++}, \tool{Objective C}, \ldots).
\tool{T2} allows users to (dis)prove
\emph{\ctl}, \emph{\textsf{Fair}-\ctl}, and \emph{\ctlstar}
specifications
via a reduction to its \emph{safety}, \emph{termination} and
\emph{nontermination} analysis techniques.
Furthermore, \emph{\ltl} 
specifications can be checked using the
automata-theoretic approach for \ltl verification~\cite{var:wol:94} via a
reduction to fair termination, which is subsumed by
\textsf{Fair}-\ctl.

In this paper we describe \tool{T2}'s capabilities and demonstrate its
effectiveness
by an experimental evaluation against competing tools.
\tool{T2} is implemented in \tool{F\#} and makes heavy
use of the \tool{Z3} SMT solver~\cite{Moura08}.
\tool{T2} runs on Windows, MacOS, and Linux.
It is available under the MIT license at \url{github.com/mmjb/T2}.

\paragraph{Related work.} We focus on tool features of \tool{T2}
and consider only related publicly released tools.
Note that, with the exception of \tool{KITTeL}~\cite{Falke11},
\tool{T2} is the only open-source termination prover and is the first open-source temporal property prover.
Similar to \tool{T2}, \tool{ARMC}~\cite{Podelski07} and
\tool{CProver}~\cite{Kroening10}, 
implement a
\tool{TERMINATOR}-style incremental reduction to safety proving. 
\tool{T2} is distinguished from these tools by its use of
lexicographic ranking functions instead of disjunctive 
termination arguments~\cite{Cook13a}.
Other termination proving tools include
\tool{FuncTion}~\cite{Urban13}, 
\tool{KITTeL}~\cite{Falke11}, and
\tool{Ultimate}~\cite{Heizmann14}, which synthesize termination
arguments, but have weak support for inferring supporting invariants in long
programs with many loops.
\tool{AProVE}~\cite{Giesl14} is a closed-source portfolio solver implementing
many successful techniques, including \tool{T2}'s methods.
We know of only one other tool able to automatically prove
\ctl properties of infinite-state programs:\footnote{We do not discuss tools that only support
finite-state systems or pushdown automata.} \tool{Q'ARMC}~\cite{Beyene13}, however \tool{Q'ARMC} does
not provide an automated front-end to its native input and requires a manual instantiation of the structure of the invariants. 
We do not know tools other than \tool{T2} that can verify
\textsf{Fair}-\ctl and \ctlstar for such programs.

\paragraph{Limitations.}
\tool{T2} only supports linear integer arithmetic fragments of \tool{C}.
An extension of \tool{T2} that handles heap program directly is presented in
\cite{Albarghouthi15}.\footnote{Alternatively, the heap-to-integer abstractions
  implemented in \tool{Thor}~\cite{Magill10} for \tool{C} or the one implemented
  in \tool{AProVE}~\cite{Giesl14} for \tool{C} and \tool{Java} can be used as a
  pre-processing step.} As in many other tools, numbers are treated as mathematical integers, not
machine integers. However, our \tool{C} front-end provides a transformation~\cite{Falke12} that
handles machine integers correctly by inserting explicit normalization steps at
possible overflows.


\vspace{-2ex}
\section{Front-end}

\ttwo improves on \tool{TERMINATOR} by supporting a native input format as well
as replacing the \tool{SLAM}-based \tool{C} interface by one based on \tool{LLVM}.

\paragraph{Native Format.}
\ttwo allows input in its internal program representation to facilitate use
from other tools.
\ttwo represents programs as graphs of program locations $\LL$
connected by transition rules with conditions and assignments to a set
of integer variables $\VV$.
The location $\ell_0 \in \LL$ is the canonical start state.
An example is shown in \rF{ex:T2-Input}(b).
We assume that variables to which we do not assign values remain
unchanged.
For precise semantics of program evaluations, we refer to
\cite{Brockschmidt13}.

\paragraph{\tool{C} via \tool{LLVM}.}
In recent years, \tool{LLVM}
has become the standard basis of program analysis tools for \tool{C}.
We have thus chosen to extend
\tool{llvm2kittel}~\cite{Falke11}, which automatically translates \tool{C}
programs into integer term rewriting systems using \tool{LLVM}, to also generate
\tool{T2}'s native format.
Our implementation
 uses the existing 
 dead code elimination, 
 constant propagation, and
 control-flow simplifications
to simplify the input program.
\rF{ex:T2-Input}(a) shows the \tool{C} program from which we generate the \tool{T2} native input
in \rF{ex:T2-Input}(b). Further details can be found in \ifisReport{the Appendix}\else{\cite{Brockschmidt15b}}\fi.

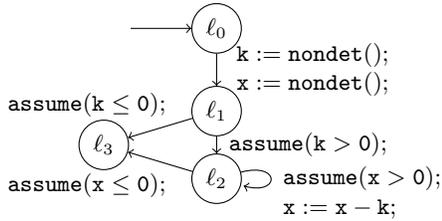
\begin{figure}[t]
\begin{tabular}{cc}
{\begin{minipage}{.36\textwidth}
  \small
\begin{alltt}
int main() \{
    int k = nondet();
    int x = nondet();
    if (k > 0)
        while (x > 0)
            x = x - k;
    return 0; \}
\end{alltt}
\end{minipage}
}
&
{\begin{minipage}{.6\textwidth}
\begin{center}
  \begin{tikzpicture}[t2]
    \node[state,color=white] (lstart) at (-1.4,0) {};
    \node[state] (l0) at (0,0)              {$\ell_0$};
    \node[state] (l0b) at ($(l0) + (0, -1.1)$) {$\ell_{1}$};
    \node[state] (l1) at ($(l0b) + (0, -.9)$) {$\ell_2$};
    \node[state] (l3) at ($(l0b) + (-1.5, -0.45)$) {$\ell_3$};

    \path[->]
    (lstart) edge (l0)
    (l0) edge 
    node[right,xshift=1mm] { $\begin{array}{l}
        \tt k := nondet();\\
        \tt x := nondet();\\
      \end{array}$
    }
    (l0b)
    (l0b) edge
    node[right]{ $\begin{array}{l}
        \tt assume(k > 0);
      \end{array}$
    }
    (l1)
    (l0b) edge
    node[above,xshift=-3em]{ $\begin{array}{l}
        \tt assume(k \leq 0);
      \end{array}$
    }
    (l3)
    
    (l1) edge[loop right]
    node[right]{ $\begin{array}{l}
        \\
        \tt assume(x > 0);\\
        \tt x := x - k;\\
      \end{array}$
    }
    (l1)
    (l1) edge
    node[below,xshift=-3em]{ $\begin{array}{l}
        \tt assume(x \leq 0);
      \end{array}$
    }
    (l3)
    ;
  \end{tikzpicture}
\end{center}
\end{minipage}}
\end{tabular}
\vspace{-3ex}
\caption{\label{ex:T2-Input} {\bf (a)} \tool{C} input program.  
{\bf (b)} \tool{T2} control-flow graph of the program in (a).}
\vspace{-5ex}
\end{figure}


\vspace{-2ex}
\section{Back-end}

In \ttwo, we have replaced the safety, termination, and non-termination procedures
implemented in \tool{TERMINATOR} by more efficient versions.
In addition, we added support for temporal-logic model checking. 

\paragraph{Proving Safety.}
To prove temporal properties, \ttwo repeatedly calls to a safety proving
procedure on instrumented programs.
For this, \tool{T2} implements the \tool{Impact}~\cite{McMillan06} safety
proving algorithm, and furthermore can use safety proving techniques\pagebreak implemented
in \tool{Z3}, {\em e.g.} generalized property directed reachability
(\tool{GPDR})~\cite{Hoder12} and \tool{Spacer}~\cite{Komuravelli14}.
For this, we convert our transition systems into sets of linear Horn
clauses with constraints in linear arithmetic, in which one predicate
$\textsf{p}_\ell$ is introduced per program location $\ell$.
For example, the transition from $\ell_2$ to $\ell_2$ in \rF{ex:T2-Input}(b) is
represented as 
 $\forall \code{x}, \code{k}, \code{x}':
   \textsf{p}_{\ell_2}(\code{x}', \code{k}) 
    \leftarrow
   \textsf{p}_{\ell_2}(\code{x}, \code{k})
    \land \code{x}' = \code{x} - \code{k}
    \land \code{x} > \code{0}$.

\begin{figure}[bt]
\scriptsize
 \center
 \begin{tikzpicture}
   \node[rectangle,draw,anchor=north west] (Preproc) at (0,0)                              {\textsf{Preproc.}};
   \node[rectangle,draw,anchor=south]      (Instr)   at ($(Preproc.north)      + (0,0.2)$) {\textsf{Instrumentation}};
   \node[rectangle,draw,anchor=west]       (Safety)  at ($(Preproc.east)       + (1.3,0)$) {\textsf{Safety}};
   \node[rectangle,draw,anchor=north west] (RFSynth) at ($(Safety.north east)  + (2,0)$)   {\textsf{RF Synth.}};
   \node[rectangle,draw,anchor=north west] (RSSynth) at ($(RFSynth.north east) + (1,0)$)   {\textsf{RS Synth.}};

   \node[anchor=north,inner sep=0.5mm,outer sep=0mm]      (Term)    at ($(Safety.south)       - (0,0.4)$) {\bfseries Termination};
   \node[anchor=north,inner sep=0.5mm,outer sep=0mm]      (Nonterm) at ($(RSSynth.south)      - (0,0.4)$) {\bfseries Nontermination};
   \node[anchor=north west] (Fail)    at ($(RSSynth.north east) + (0.9,0)$) {\bfseries Fail};

   \path[->, thick]
     ($(Instr.north) + (0,0.2)$) edge (Instr)
     ($(Preproc.north) + (0,0.2)$) edge (Preproc)
     (Preproc) edge             node[above] {Simplif.}   (Safety)
     (Safety)  edge             node[above] {Counterex.} (RFSynth)
     (Safety)  edge             node[left]  {Safe}       (Term)
     (RFSynth) edge             node[above] {Fail}       (RSSynth)
     (RFSynth) edge[bend right=25] node[above] {Refine}     (Safety)
     (RSSynth) edge             node[above] {Fail}       (Fail)
     (RSSynth) edge             node[left]  {Succ.}      (Nonterm)
   ;
 \end{tikzpicture}
 \vspace*{-3ex}
 \caption{Flowchart of the \tool{T2} termination proving procedure}
 \label{fig:t2-term-flowchart}
 \vspace{-2em}
\end{figure}
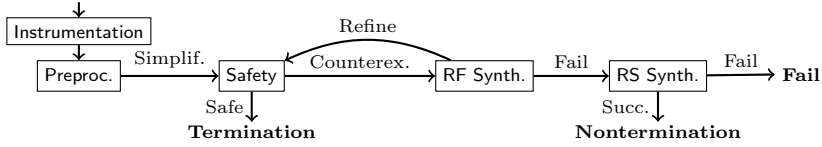

\paragraph{Proving Termination.}
A schematic overview of our termination proving procedure is displayed in
\rF{fig:t2-term-flowchart}.
In the initial \textsf{Instrumentation} phase (described in
\cite{Brockschmidt13}), the input program is modified so that a termination
proof can be constructed by a sequence of alternating safety queries and rank
function synthesis steps.
This reduces the check of a speculated (possibly lexicographic) rank function
$f$ for a loop to asserting that the value of $f$ after one loop iteration is
smaller than before that iteration.
If the speculated termination argument is insufficient, our \textsf{Safety}
check fails, and the termination argument is refined using the found
counterexample in \textsf{RF Synth}.
We follow the strategy presented in \cite{Cook13a} to construct a
lexicographic termination argument, extending a standard linear rank function
synthesis procedure~\cite{Podelski04b},\footnote{\tool{T2} can optionally also
  synthesize disjunctive termination arguments~\cite{Podelski04a} as implemented
  in the original \tool{TERMINATOR}~\cite{Cook06a}.}
implemented as constraint solving via \tool{Z3}.
The overall procedure is independent of the used safety prover and
rank function synthesis.

In our \textsf{Preprocessing} phase, a number of standard program analysis
techniques are used to simplify the remaining proof.
Most prominently, this includes the termination proving pre-processing technique 
presented in \cite{Brockschmidt13} to remove loop transitions that we can
directly prove terminating, without needing further supporting invariants.
In our termination benchmarks, about 80\% of program loops ({\em e.g.} encodings
of \texttt{for i in 1 .. n do}-style loops) are eliminated at this stage.

\paragraph{Disproving Termination.}
When \tool{T2} cannot refine a termination argument based on a given
counterexample, it tries to prove existence of a recurrent
set~\cite{Gupta08} witnessing non-termination in the \textsf{RS Synth.}
step.
A recurrent set $S$ is a set of program states whose execution can
eventually lead back to a state from $S$.
\tool{T2} uses a variation of the techniques from
\cite{Brockschmidt12a}, restricted to only take a counterexample
execution into account and implemented as constraint solving via \tool{Z3}.


\paragraph{Proving \ctl.}
\label{sect:CTL}
\ctl subsumes reasoning about safety, termination, and nontermination, in addition
to all state-based properties.
\ttwo implements the bottom-up strategy for \ctl verification
from~\cite{Cook14}.
Given a \ctl property $\varphi$, \ttwo first computes quantifier-free
preconditions $\textsf{precond}_i$ for the subformulas of $\varphi$,
and then verifies the formula obtained from $\varphi$ by replacing the
subformulas by their preconditions.
Property preconditions are computed using a
counterexample-guided strategy where 
several preconditions for each location are computed simultaneously
through the natural decomposition of
the counterexample's state space. 

\paragraph{Proving \textsf{Fair}-\ctl.} 
\tool{T2} implements the approach for verification of \ctl with
fairness as presented in~\cite{coo:khl:pit:15}.
This method reduces \textsf{Fair}-\ctl to fairness-free \ctl using
prophecy variables to encode a partition of fair from unfair paths.
Although \ctl can express a system's interaction with inputs and nondeterminism, 
which linear-time temporal logics (\ltl) are inadequate to express, it cannot model trace-based assumptions about the environment in sequential and concurrent settings (e.g. schedulers) that \ltl can express. Fairness allows us to bridge said gap between linear-time and branching-time reasoning, in addition to allowing us to employ the automata-theoretic technique for \ltl verification~\cite{var:wol:94} in \tool{T2}. 

\paragraph{Proving \ctlstar.} 
Finally, \tool{T2} is the sole tool which supports the verification of \ctlstar
properties of infinite-state programs as presented in~\cite{CookKP15}.
A precondition synthesis strategy is used with a program transformation
that trades nondeterminism in the transition 
relation for nondeterminism explicit in variables predicting future
outcomes when necessary. Note that \textsf{Fair}-\ctl disallows
the arbitrary interplay between linear-time and branching-time
operators beyond the scope of fairness. For example, a property stating that
``along {\em some} future an event occurs {\em
  infinitely often}" cannot be expressed in either \ltl, \ctl nor \textsf{Fair}-\ctl,
yet it is crucial when expressing ``possibility" properties, such as the viability of a system, stating that every reachable state can spawn a fair computation. 
Contrarily, \ctlstar is capable of expressing \ctl, \ltl, \textsf{Fair}-\ctl, and
the aforementioned property. Additionally, \ctlstar allows us to express existential system stabilization, stating that an event can eventually become true and stay true from every reachable state.
Note that for properties expressible in \textsf{Fair}-\ctl, our
\textsf{Fair}-\ctl prover is relatively (to safety and termination
subprocedures) complete, whereas our \ctlstar prover is incomplete.


\vspace{-2ex}
\section{Experimental Evaluation \& Future Work}

We demonstrate \tool{T2}'s effectiveness compared to competing tools.
We do not know of other tools supporting {\textsf Fair}-\ctl and
\ctlstar for infinite-state systems, thus we do not present such experiments
and instead refer to \cite{coo:khl:pit:15} and \cite{CookKP15}.
Note that \tool{T2}'s performance has significantly improved since then through
improvements in our back-end (e.g. by using \tool{Spacer} instead of \tool{Impact}).
We refer to \ifisReport{to the Appendix}\else{\cite{Brockschmidt15b}}\fi{}
for a detailed discussion of the properties and programs
that these logics allowed us to verify.

\paragraph{Termination Experiments.}
We compare \tool{T2} as termination prover with the participants of the
Termination Competition 2014 and 2015 using the collection of 1222
termination proving benchmarks used at the Termination Competition
2015 for integer transition systems.
These benchmarks include manually crafted programs from the literature on
termination proving, as well as many examples obtained from automatic
translations from programs in higher languages such as \tool{Java} ({\em e.g.}
from \code{java.util.HashSet}) or \tool{C} ({\em e.g.} reduced versions of
Windows kernel drivers).
The experiments were performed on the StarExec platform with a timeout of 300
seconds.
Our version of \tool{T2} uses the \tool{GPDR} implementation in \tool{Z3} as
safety prover.
Furthermore, we also consider three further versions of \tool{T2}, using the
three different supported safety provers.
For these configurations, we use no termination proving pre-processing
(\textsf{NoP}) step and only use our safety proving-based strategy, to better
evaluate the effect of different safety back-ends.
The overall number of solved instances and average runtimes are displayed in
\rF{fig:Term-Evaluation}(a), and a 
detailed comparison of \tool{AProVE} and \tool{T2-GPDR} is shown in
\rF{fig:Term-Evaluation}(b).\footnote{All experimental data can be viewed on
  \url{https://www.starexec.org/starexec/secure/details/job.jsp?id=11121}.}
All provers are assumed to be sound, and no provers returned conflicting
results.

\begin{figure}[t]
  \begin{minipage}[t]{.57\linewidth}
  \vspace{-4cm}
  \center
  \begin{tabular}{@{}l@{\ }c@{\ }c@{\ }c@{\ }c@{}}
    \toprule
    Tool                   & \textbf{Term} &  \textbf{Nonterm} & \textbf{Fail} & \textbf{Avg.} (s)\\
    \midrule
    \tool{AProVE}          &        641 &         393 &        188 &    49.1   \\
    \tool{CppInv}          &        566 &         374 &        282 &    65.5   \\
    \tool{Ctrl}            &        445 &\phantom{00}0&        777 &    80.0   \\
    \midrule
    \tool{T2-GPDR}         &        627 &         442 &        153 &    23.6   \\
    \midrule
    \tool{T2-GPDR-NoP}     &        589 &         438 &        195 &    31.4   \\
    \tool{T2-Spacer-NoP}   &        591 &         429 &        202 &    33.5   \\
    \tool{T2-Impact-NoP}   &        529 &         452 &        241 &    37.2   \\
    \bottomrule
  \end{tabular}
  \end{minipage}
  \begin{minipage}[t]{.4\linewidth}
  \begin{tikzpicture}
    \begin{axis}[
        scatter/classes={
          YY={mark=o,draw=green},%
          YM={mark=square,draw=green},%
          MY={mark=triangle,draw=green},%
          NN={mark=o,draw=blue},%
          NM={mark=square,draw=blue},%
          MN={mark=triangle,draw=blue},%
          MM={mark=o,draw=gray},
          NY={mark=o*,draw=red},
          YN={mark=o*,draw=red}},
        width=4.75cm,
        height=4.75cm,
        xlabel=\textsf{T2-GPDR} (s), ylabel shift=-1.5ex,
        ylabel=\textsf{AProVE} (s), xlabel shift=-0.5ex,
        ymin=0.5,ymax=300, xmin=0.5,xmax=300,
        xtick={0.5,1,2,3,4,5,6,7,8,9,10,20,30,40,50,60,120,180,240,300},
        xticklabels={0.5,1,,,,5,,,,,10,,30,,,60,,,,300},
        ytick={0.5,1,2,3,4,5,6,7,8,9,10,20,30,40,50,60,120,180,240,300},
        yticklabels={0.5,1,,,,5,,,,,10,,30,,,60,120,,,300},
        xmode=log,ymode=log,
        ]
    \addplot+[scatter,only marks,point meta=explicit symbolic]
             table[meta=label] {experiments/GPDR-v-AProVE.data};
    \addplot [black,dashed] plot coordinates {(0.5,0.5) (300,300)}; 
    \addplot [black] plot coordinates {(0.5,300) (300,300)}; 
    \addplot [black] plot coordinates {(300,0.5) (300,300)}; 
    \end{axis}
  \end{tikzpicture}
  \end{minipage}
  \\
  \centerline{
    \begin{minipage}{0.6\textwidth}
      \vspace{-3ex}
      \hspace{1cm}(a)\hspace{2.1in}(b)
    \end{minipage}}
  \vspace{-6ex}
  \caption{\small Termination evaluation results.
    (a) Overview table. (b) Comparison of \tool{T2} and \tool{AProVE}.
    \textcolor{green}{Green} (resp. \textcolor{blue}{blue}) marks correspond to
    terminating (resp. non-terminating) examples,
    and \textcolor{gray}{gray} marks examples on which both provers failed.
    A $\square$ (resp. a $\triangle$) indicates an example in which only
    \tool{T2} (resp. \tool{AProVE}) succeeded, and $\circ$ indicates an example on
    which both provers return the same result.}
  \label{fig:Term-Evaluation}
  \vspace{-4ex}
\end{figure}
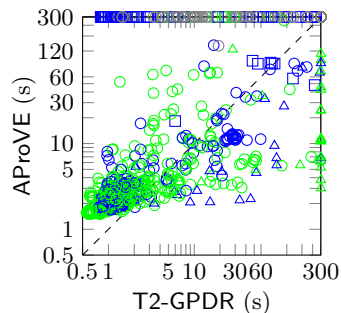

The results show that \tool{T2}'s simple architecture competes well with the
portfolio approach implemented in \tool{AProVE} (which subsumes \tool{T2}'s
techniques), and is more effective than other tools.
Comparing the different safety proving back-ends of \tool{T2} shows that our
\tool{F\#} implementation of \tool{Impact} is nearly as efficient as the
optimized \tool{C++} implementations of \tool{GPDR} and \tool{Spacer}.
The different exploration strategies of our safety provers yield different
counterexamples, leading to differences in the resulting (non)termination proofs.
The impact of our pre-processing technique is visible when comparing
\tool{T2-GPDR} and \tool{T2-GPDR-NoP}.

\paragraph{\ctl Experiments.}

\begin{wrapfigure}[7]{r}{0.28\textwidth}
  \vspace{-1.2cm}
  \hspace*{-2.5ex}
  \begin{tikzpicture}
    \begin{axis}[
        width=4cm,
        height=4cm,
        xlabel={\footnotesize\textsf{T2} (s)}, xlabel shift=-1.3ex,
        ylabel={\footnotesize\textsf{QARMC} (s)}, ylabel shift=-1.7ex,
        ymin=0.5,ymax=100, xmin=0.5,xmax=100,
        xtick={0.5,1,2,3,4,5,6,7,8,9,10,20,30,40,50,60,100},
        xticklabels={\footnotesize0.5,\footnotesize1,,,,\footnotesize5,,,,,\footnotesize10,,\footnotesize30,,,,\footnotesize100},
        ytick={0.5,1,2,3,4,5,6,7,8,9,10,20,30,40,50,60,100},
        yticklabels={\footnotesize0.5,\footnotesize1,,,,\footnotesize5,,,,,\footnotesize10,,\footnotesize30,,,,\footnotesize100},
        xmode=log,ymode=log,
        ]
    \addplot [black,dashed] plot coordinates {(0.5,0.5) (300,300)}; 
    \addplot [black] plot coordinates {(0.5,300) (300,300)}; 
    \addplot [black] plot coordinates {(300,0.5) (300,300)}; 
    \addplot [only marks,mark=x,draw=black] table {experiments/T2-v-QARMC.data};
    \end{axis}
  \end{tikzpicture}
\end{wrapfigure}
We evaluate \tool{T2}'s \ctl verification techniques against the only other
available tool, \tool{Q'ARMC}~\cite{Beyene13} on the 56 benchmarks from its
evaluation.
These benchmarks
are drawn from the I/O subsystem of the Windows OS kernel, the back-end
infrastructure of the PostgreSQL database server, and the SoftUpdates
patch system. They can be found at \url{http://www.cims.nyu.edu/~ejk/ctl/}.
The tools were executed on a Core i7 950 CPU with a timeout of 100
seconds.
Both tools are able to successfully verify all examples. \tool{T2} needs 2.7
seconds on average, whereas \tool{Q'ARMC} takes 3.6 seconds.
The scatterplot above compares proof times on individual
examples.


\paragraph{Future work.}
We wish to integrate and improve techniques for
conditional termination, which will improve the strength of
our property verification.
We also intend to support reasoning about the heap, recursion, and concurrency
in \tool{T2}.


\vspace{-3ex}
\bibliographystyle{plain}
\bibliography{strings-short,references,crossrefs-short}

\begin{thebibliography}{10}

\bibitem{Albarghouthi15}
A.~Albarghouthi, J.~Berdine, B.~Cook, and Z.~Kincaid.
\newblock Spatial interpolants.
\newblock In {\em ESOP'15}.

\bibitem{Beyene13}
T.~A. Beyene, C.~Popeea, and A.~Rybalchenko.
\newblock Solving existentially quantified horn clauses.
\newblock In {\em CAV'13}.

\bibitem{Brockschmidt13}
M.~Brockschmidt, B.~Cook, and C.~Fuhs.
\newblock Better termination proving through cooperation.
\newblock In {\em CAV'13}.

\bibitem{Brockschmidt12a}
M.~Brockschmidt, T.~Str\"oder, C.~Otto, and J{\"u}rgen Giesl.
\newblock Automated detection of non-termination and
  \texttt{NullPointerExceptions} for \tool{Java Bytecode}.
\newblock In {\em FOVEOOS'11}.

\bibitem{coo:khl:pit:15}
B.~Cook, H.~Khlaaf, and N.~Piterman.
\newblock Fairness for infinite-state systems.
\newblock In {\em TACAS'15}.

\bibitem{Cook14}
B.~Cook, H.~Khlaaf, and N.~Piterman.
\newblock Faster temporal reasoning for infinite-state programs.
\newblock In {\em FMCAD'14}.

\bibitem{CookKP15}
B.~Cook, H.~Khlaaf, and N.~Piterman.
\newblock On automation of \ctlstar verification for infinite-state systems.
\newblock In {\em CAV'15}.

\bibitem{Cook06a}
B.~Cook, A.~Podelski, and A.~Rybalchenko.
\newblock Termination proofs for systems code.
\newblock In {\em PLDI'06}.

\bibitem{Cook13a}
B.~Cook, A.~See, and F.~Zuleger.
\newblock Ramsey vs. lexicographic termination proving.
\newblock In {\em TACAS'13}.

\bibitem{Moura08}
L.~de~Moura and N.~Bj{\o}rner.
\newblock \tool{Z3}: An efficient {SMT} solver.
\newblock In {\em TACAS'08}.

\bibitem{Falke12}
S.~Falke, D.~Kapur, and C.~Sinz.
\newblock Termination analysis of imperative programs using bitvector
  arithmetic.
\newblock In {\em VSTTE'12}.

\bibitem{Falke11}
S.~Falke, D.~Kapur, and C.~Sinz.
\newblock Termination analysis of \tool{C} programs using compiler intermediate
  languages.
\newblock In {\em RTA'11}.

\bibitem{Giesl14}
J.~Giesl, M.~Brockschmidt, F.~Emmes, F.~Frohn, C.~Fuhs, C.~Otto,
  M.~Pl{\"{u}}cker, P.~Schneider{-}Kamp, T.~Str{\"{o}}der, S.~Swiderski, and
  R.~Thiemann.
\newblock Proving termination of programs automatically with \tool{AProVE}.
\newblock In {\em IJCAR'14}.

\bibitem{Gupta08}
A.~Gupta, T.~Henzinger, R.~Majumdar, A.~Rybalchenko, and R.~Xu.
\newblock Proving non-termination.
\newblock In {\em POPL'08}.

\bibitem{Heizmann14}
M.~Heizmann, J.~Hoenicke, and A.~Podelski.
\newblock Termination analysis by learning terminating programs.
\newblock In {\em CAV'14}.

\bibitem{Hoder12}
K.~Hoder and N.~Bj{\o}rner.
\newblock Generalized property directed reachability.
\newblock In {\em SAT'12}.

\bibitem{Komuravelli14}
A.~Komuravelli, A.~Gurfinkel, and S.~Chaki.
\newblock {SMT}-based model checking for recursive programs.
\newblock In {\em CAV'14}.

\bibitem{Kroening10}
D.~Kroening, N.~Sharygina, A.~Tsitovich, and C.~Wintersteiger.
\newblock Termination analysis with compositional transition invariants.
\newblock In {\em CAV'10}.

\bibitem{Magill10}
S.~Magill, M.~Tsai, P.~Lee, and Y.~Tsay.
\newblock Automatic numeric abstractions for heap-manipulating programs.
\newblock In {\em POPL'10}.

\bibitem{McMillan06}
K.~McMillan.
\newblock Lazy abstraction with interpolants.
\newblock In {\em CAV'06}.

\bibitem{Podelski04b}
A.~Podelski and A.~Rybalchenko.
\newblock A complete method for the synthesis of linear ranking functions.
\newblock In {\em VMCAI'04}.

\bibitem{Podelski07}
A.~Podelski and A.~Rybalchenko.
\newblock \tool{ARMC}: The logical choice for software model checking with
  abstraction refinement.
\newblock In {\em PADL'07}.

\bibitem{Podelski04a}
A.~Podelski and A.~Rybalchenko.
\newblock Transition invariants.
\newblock In {\em LICS'04}.

\bibitem{Urban13}
C.~Urban.
\newblock The abstract domain of segmented ranking functions.
\newblock In {\em SAS'13}.

\bibitem{var:wol:94}
M.Y. Vardi and P.~Wolper.
\newblock Reasoning about infinite computations.
\newblock {\em Inf. Comput.}, 115(1):1--37, 1994.

\end{thebibliography}

\ifisReport
  \newpage
  \appendix
  \noindent
    {\bf \large Appendix.}
\bigskip
    
\noindent
This appendix contains several examples on how to use \tool{T2}.
For the ease of this demonstration, we include easy to follow programs alongside
corresponding simple properties. Additional examples of \tool{T2} operating on realistic programs 
with expressive properties are available in the papers relating to the respective technical
results~\cite{Cook13a,Cook14,coo:khl:pit:15}.
Installation instructions for \tool{T2}, additional runtime options, and an overview of the program source
code can be found alongside its source in
\url{https://github.com/mmjb/T2/blob/master/README.txt}.

\section{Front-end Pre-processing via LLVM}
\label{llvm}

\begin{figure*}
\vspace{-3ex}
\begin{tabular}{cc}
{\begin{minipage}{0.35\textwidth}
  \small
\begin{alltt}
define i32 @main() #0 \{
main_bb0:
  \%"0" = call i32 (...)* @nondet()
  \%"1" = call i32 (...)* @nondet()
  \%"2" = icmp sgt i32 \%"0", 0
  br i1 \%"2", label \%main_bb1, 
      label \%main_bb3

main_bb1:
  \%x = phi i32 [ \%"4", \%main_bb2],
      [ \%"1", \%main_bb0 ]
  \%"3" = icmp sgt i32 \%x, 0
  br i1 \%"3", label \%main_bb2,
      label \%main_bb3

main_bb2:
  \%"4" = sub nsw i32 \%x, \%"0"
  br label \%main_bb1

main_bb3:
  ret i32 0
\}
\end{alltt}
\end{minipage}
}
&
\begin{tabular}{cc}
 \null\hspace{0.5in}
  \begin{minipage}{.26\textwidth}
  \small
  \begin{alltt}
START: main_bb0;

FROM: main_bb0;
  v0 := nondet();
  v1 := nondet();
  x := v1;
TO: main_bb0_end;

FROM: main_bb0_end;
  assume(v0 > 0);
TO: main_bb1;

FROM: main_bb0_end;
  assume(v0 <= 0);
TO: main_bb3;
\end{alltt}
\end{minipage}
&
\begin{minipage}{.3\textwidth}
\begin{alltt}


FROM: main_bb1;
  assume(x > 0);
TO: main_bb2;

FROM: main_bb1;
  assume(x <= 0);
TO: main_bb3;

FROM: main_bb2;
  v4 := x - v0;
  x := v4;
TO: main_bb1;

FROM: main_bb3;
TO: main_bb3;
  \end{alltt}
 \end{minipage}
 \end{tabular}
 \end{tabular}
\\
\centerline
{\begin{minipage}{0.5\textwidth}
\vspace{-1ex}
(a)\hspace{2in}(b)
\end{minipage}}
\\
\vspace{-3ex}
\caption{\label{ex:T2} {\bf (a)} Compiled LLMV-IR post \tool{llvm2kittel} optimizations 
corresponding to \rF{ex:T2-Input}(a).  
{\bf (b)} \tool{T2} input file corresponding to \rF{ex:T2-Input}(b), generated from(a).}
\vspace{-3ex}
\end{figure*}

 Our LLVM front-end builds upon and extends
\tool{llvm2kittel}~\cite{Falke11}. Our version of
  \textsf{llvm2kittel} tailored for \textsf{T2} can be found at
  \url{https://github.com/hkhlaaf/llvm2kittel}.
\tool{llvm2kittel} provides multiple optimizations that are helpful for our
transformation into the native \ttwo file format, as it performs function
inlining, dead code elimination, constant propagation, and control-flow
simplification. 
Below we provide a very basic notion of how the \tool{LLVM} intermediate
representation (\tool{LLVM-IR}) corresponds to the \tool{T2} format.

The \tool{LLVM-IR} generated by \code{clang} for our example \tool{C} program
from \rF{ex:T2-Input}(a) is shown in \rF{ex:T2}(a).
The \tool{T2} input file generated from this by our \tool{llvm2kittel} front-end
is displayed in \rF{ex:T2}(b).
In our translation, basic blocks in the \tool{LLVM-IR} (\texttt{main\_bb0},
\texttt{main\_bb1}, \ldots) are translated as transition rules labeled with
corresponding arithmetic instructions.
These instructions are trivially obtained from the \tool{LLVM-IR}, but
all heap memory reads are implemented as \code{\bf nondet()}, and heap writes
are dropped.

A basic block's entry point is represented by a location of the same name, i.e.,
a transition to the location \code{main\_bb2} corresponds to entering the basic
block \code{main\_bb2}.
The targets of the generated transitions are extracted from the \texttt{br}
(``branch'') instructions.
Sequences of \texttt{phi} instructions at the beginning of a basic block $b$,
which are needed for \tool{LLVM-IR}'s single static assignment syntax, are
encoded on the transitions leading to $b$.
For example, in \rF{ex:T2}(a), the basic block \texttt{main\_bb0} contains a
sequence of instructions before a \texttt{br} instruction determines
whether to branch to \texttt{main\_bb1} or \texttt{main\_bb3}, depending on the
value of \texttt{\%0}. 
This is reflected in \rF{ex:T2}(b) in the first column, where the comparison of
the value \texttt{\%0} (\texttt{v0} in the \ttwo file), is done from the 
\texttt{main\_bb0\_end} node.
If \texttt{v0 > 0} we transition to the \texttt{main\_bb1} node, otherwise we
transition  to the \texttt{main\_bb3} node.

Using our version of \tool{llvm2kittel} as a front-end, we now show how it can be used to
generate native \ttwo files from \tool{C} programs. Assume that the \tool{C}
program from \rF{ex:T2-Input}(a) is stored as \code{ex0.c}. We generate a \ttwo 
native input file as follows:
\begin{quote}
  \vspace{-1ex}
  \begin{alltt}
    \small
\$ clang -Wall -Wextra -c -emit-llvm -O0 ex0.c -o ex0.bc
\$ ./llvm2kittel --eager-inline --t2 ex0.bc > ex0.t2
  \end{alltt}
  \vspace{-1ex}
\end{quote}

\section{\tool{T2} as Termination Prover}
\subsection{Native Input}
We first demonstrate using \tool{T2} to prove termination of the example from
\rF{ex:T2-Input}, whose textual representation is displayed in \rF{ex:T2}
Assume that the example is saved as file \code{ex0.t2}. Then, 
the most simple \tool{T2} call looks like this:
\begin{quote}
  \vspace{-2ex}
\begin{alltt}
\small
\$ ./T2 -termination -input\_t2 ex0.t2
Termination proof succeeded
\end{alltt}
  \vspace{-1ex}
\end{quote}
To obtain more information about the termination argument, \tool{T2} provides
the \code{-print\_proof} option:
\begin{quote}
  \vspace{-2ex}
\begin{alltt}
\small
\$ ./T2 -termination -input\_t2 ex0.t2 -print\_proof
Termination proof succeeded
Used the following cutpoint-specific lexicographic rank functions:
 * For cutpoint 7, used the following rank functions/bounds (in descending priority order):
    - RF x, bound 1
\end{alltt}
  \vspace{-1ex}
\end{quote}
We see that the proof was done using a (one-element) lexicographic rank function.
However, this output is hard to connect to the input program, which had no
location \code{7}.\footnote{The reason for the location number is that \tool{T2}
  stores 
  locations as integers, but also allows strings to identify locations in the
  input (e.g. ``\code{START: start;}''), and thus renumbers all locations on
  parsing the input file.}
To understand the connection better, \tool{T2} allows to output all intermediate
program representations as \tool{DOT} graphs:
\begin{quote}
  \vspace{-1ex}
\begin{alltt}
\small
\$ ./T2 -termination -input\_t2 ex0.t2 -dottify\_input\_pgms
Created input.dot
Created input__instrumented.dot
Created input__instrumented_cleaned.dot
Created input__instrumented_lex_RF.dot
Termination proof succeeded
\end{alltt}
  \vspace{-1ex}
\end{quote}
In general, \code{input.dot} corresponds to the parsed program (with renamed
locations and numbered transitions), \code{input\_\_instrumented.dot} shows it
after instrumentation for a termination proof, and
\code{input\_\_instrumented\_cleaned.dot} is the program after the initial
\textsf{Preprocessing} step (cf. \rF{fig:t2-term-flowchart}).
A rendering of the \code{input\_\_instrumented.dot} file is shown in
\rF{fig:ex0-input__instrumented.dot}. Location are circular nodes in the graph,
and the labels ``loc\_$i$'' indicate which node corresponds to location $i$ in
the input program.

\begin{figure}
  \includegraphics[scale=.25]{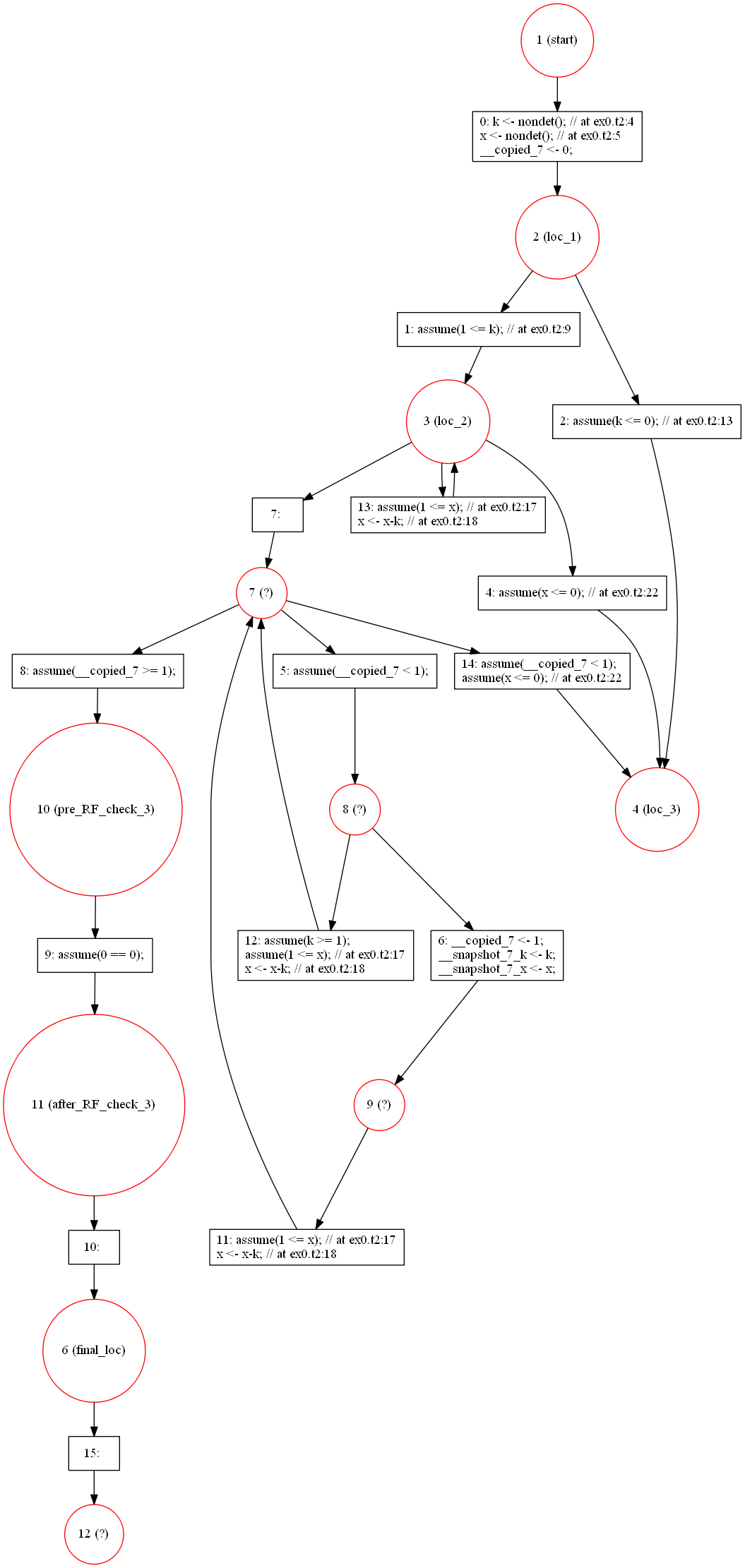}
  \caption{\label{fig:ex0-input__instrumented.dot}CFG for \rF{ex:T2-Input} after
  instrumentation for termination}
\end{figure}


\subsection{\tool{Java} Input}

\begin{figure}[t]
 \begin{minipage}[t]{.5\linewidth}
  \begin{alltt}
\small
public class Ex1 \{
  public static void 
   main(String... args) \{
    int n = args.length;
    int m = 2 * (n + 1);
    while (n > 0) \{
      if (m <= 0) \{
        n++; m++;
      \} else \{
        n--; m--; \}\}\}\}
  \end{alltt}
 \end{minipage}
 \begin{minipage}[t]{.5\linewidth}
  \begin{alltt}
\small
public class Ex2 \{
  private Ex2 next = null;
  public Ex2(Ex2 n) \{ next = n; \}
  public static void 
   main(String... args) \{
    int n = args.length;
    Ex2 list = null;
    while (--n > 0)
      list = new Ex2(list);
    int length = 0;
    while (list != null) \{
      length++;
      list = list.next; \}\}\}
  \end{alltt}
 \end{minipage}
 \vspace{-2ex}
 \caption{\label{fig:java-ex}Two \tool{Java} example programs}
 \vspace{-4ex}
\end{figure}
Using \tool{AProVE}~\cite{Giesl14} as frontend, \tool{T2} can be used to prove
termination of \tool{Java} programs. As an example, consider the small
\tool{Java} program \code{Ex1} in \rF{fig:java-ex}. As \tool{AProVE} only
supports reading \code{JAR} files (i.e., compiled \tool{Java} code), we will
assume that the example was compiled to \code{Ex1.jar}, and that \tool{AProVE}
is available as \code{aprove.jar}.\footnote{This is downloadable from
  \url{http://aprove.informatik.rwth-aachen.de/}.}
We can then use \tool{AProVE} to obtain a \tool{T2} file, which we then prove
terminating:
\begin{quote}
  \vspace{-2ex}
  \begin{alltt}
    \small
\$ java -cp aprove.jar aprove.CommandLineInterface.JBCFrontendMain --t2 yes Ex1.jar
Dumped to ./Ex1.jar-obl-8.t2
\$ ./T2 -termination -input_t2 Ex1.jar-obl-8.t2
Termination proof succeeded
  \end{alltt}
  \vspace{-2ex}
\end{quote}
We note that \tool{AProVE} cannot prove this example terminating on its own, as
it cannot infer the needed invariant $\code{n} < \code{m}$.
\tool{AProVE} also supports heap-manipulating programs, and can translate
these into integer transition systems, which can then be handled by
\tool{T2}. As example, consider the example program \code{Ex2} in
\rF{fig:java-ex}, in which a list is first constructed and its length is
subsequently computed. We can prove termination of it as follows:
\begin{quote}
  \vspace{-2ex}
  \begin{alltt}
    \small
\$ java -cp aprove.jar aprove.CommandLineInterface.JBCFrontendMain --t2 yes Ex2.jar
Dumped to ./Ex2.jar-obl-9.t2
\$ ./T2 -termination -input_t2 Ex2.jar-obl-9.t2
Termination proof succeeded
  \end{alltt}
  \vspace{-2ex}
\end{quote}


\section{Temporal Property Verification}
\subsection{\tool{T2} as a \ctl Prover}

In this section, we demonstrate how one can verify \tool{C} programs
using the \ctl option in \textsf{T2}. In the following demonstration, we will
show how we can verify the property $\mathsf{EFAG}\;\texttt{x} \leq 0$.
As demonstrated above, we use \tool{llvm2kittel} to generate a \ttwo
input file for the program from \rF{ex:T2-Input}(a), stored as \code{ex0.t2}.
Note that the LLVM compilation process may slightly 
modify program variable names. Thus, the variables used to specify the \ctl property must be changed accordingly
as well. We now run \textsf{T2} as follows:
\begin{quote}
\small
\begin{alltt}
\$ ./T2 -input_t2 ctl-ex.t2 -CTL "[EF]([AG](x <= 0))"
T2 program prover/analysis tool.
Temporal proof succeeded
\end{alltt}
\end{quote}
One can additionally specify the \texttt{-print\_proof} option, which
outputs the location-specific preconditions generated for each
sub-formula. The precondition is a tuple with the first argument being
a program location, and the second being the precondition. That is, a
precondition $a_\varphi$ for a \ctl sub-formula $\varphi$ takes the
form $\bigwedge_{i}(\textsf{pc}= i \Rightarrow a_{\textsf{pc}_i})$
where $i$ denotes elements of the program locations.

\subsection{\tool{T2} as a \textsf{Fair}-\ctl and \ctlstar Prover}

Below we show properties which can be expressed in \textsf{Fair}-\ctl and \ctlstar, but not
\ctl nor \ltl. We write these properties in \ctlstar, a superset of \ctl and
\ltl. 

\paragraph{Properties expressible in \textsf{Fair}-\ctl.} 
For brevity, when expressing \textsf{Fair}-\ctl properties we write
$\Omega$ for $\G\F p{\rightarrow}\G\F q$. 
A state property is indicated by $\varphi$ and $p$ and $q$
are subsets of program states, constituting our fairness
requirement (infinitely often $p$ implies infinitely often $q$). 

The property  $\E  [\Omega\wedge \G\varphi ]$ generalizes fair
non-termination, that is, there exists an infinite fair computation
all of whose states satisfy the property $\varphi$.
The property $\A \big [\Omega\rightarrow \G
[\varphi_1\rightarrow \A (\Omega \rightarrow \F\varphi_2 )  ] \big ]$
indicates that on every fair path, every $\varphi_1$
state is later followed by a $\varphi_2$ state.
In ~\cite{coo:khl:pit:15}, we verify said property for a Windows device driver,
indicating that a lock will always eventually be released
in the case
that a call to a lock occurs, 
provided that whenever we continue to
call a Windows API repeatedly, it will eventually return a desired
value (fairness). 
Similarly, $\A\big [
\Omega \rightarrow \G
 [\varphi_1 \rightarrow \A (\Omega \rightarrow \F\E (\Omega \wedge \G\varphi_2 )) ] \big ]
$ dictates that on every fair path whenever a
$\varphi_1$ state is reached, on all possible futures there is
a state which is a possible fair future and $\varphi_2$ is always
satisfied. For example, one may wish to verify that there will be a possible active fair
continuation of a server, and that it will continue to effectively serve
if sockets are successfully opened. Below we demonstrate how we can verify our Bakery algorithm benchmark from~\cite{coo:khl:pit:15} with a \ctl property and a fairness constraint $\Omega$ for $\G\F p{\rightarrow}\G\F q$:
\begin{quote}
\small
\begin{alltt}
\$ ./T2 -input_t2 test/bakery.t2 
         -CTL "[AG](NONCRITICAL <= 0 || ([AF](CRITICAL > 0)))"
         -fairness "(P == 1, Q == 1)"
T2 program prover/analysis tool.
Temporal proof succeeded
\end{alltt}
\vspace{-1ex}
\end{quote}

\paragraph{Properties expressible in \ctlstar.} 
Below are properties that can only be afforded by the extra
expressive power of \ctlstar, which subsumes \textsf{Fair}-\ctl. These liveness properties are 
utilized in ~\cite{CookKP15} to verify
systems such as Windows kernel APIs that
acquire resources and APIs that release resources.

The property $\E\F\G(\neg x \wedge (\E\G\F\; x))$ conveys
the divergence of paths. That is, there is a path in which a system
stabilizes to $\neg x$, but every point on said path has a
diverging path in which $x$ holds infinitely often. 
This property is not expressible in \ctl or in \ltl, yet 
is crucial when expressing the existence of fair paths
spawning from every reachable state in a system.
In \ctl, one can only examine sets of states, disallowing us
to convey properties regarding paths. 
In \ltl, one cannot approximate a solution by trying to {\em disprove}
either $\F\G\;\neg x$ or $\G\F\;x$, as one cannot characterize these proofs
within a path quantifier. 

Another \ctlstar property $\A\G\big[ (\E\G \;\neg x ) \vee
  (\E\F\G\;y)\big]$ dictates that from every state of a
program, there exists either a computation in which $x$ never
holds or a computation in which $y$ eventually always holds.  
The linear time property $\G(\F x \rightarrow \F\G\;y)$
is significantly stricter as it requires that on every computation
either the first disjunct or the second disjunct hold. Finally, the
property $\E\F\G\big[(x \vee (\A\F\;\neg y))\big]$
asserts that there exists a computation in which whenever $x$
does not hold, all possible futures of a system lead to the
falsification of $y$. This assertion is impossible to express
in \ltl. Below we demonstrate how we can verify one of our benchmarks from~\cite{CookKP15}:

\begin{quote}
\small
\begin{alltt}
\$ ./T2 -input_t2 1394-succeed-2.t2
         -ctlstar "E F(G (((keA <= 0) || (E F (keR == 1)))))
T2 program prover/analysis tool.
Temporal proof succeeded
\end{alltt}
\end{quote}

  \section{\tool{T2} options}

\tool{T2} provides a \code{--help} command line switch. However, the following
switches are noteworthy:
\begin{itemize}
 \item \code{--log} turns on live logging, so that \tool{T2} reports every
   attempted proof step in detail (e.g., expansion of leaves in the \tool{Impact} 
   safety procedure, found counterexamples, program refinements, ...).
 \item \code{--safety\_implementation} allows to pick the used back-end safety
   solver. Currently, this supports the internal \code{impact}, and \tool{Z3}'s
   \code{spacer} (the default) and \code{pdr}. There is also an experimental
   mode that runs \code{spacer} and \code{bmc} (bounded model checking) in
   parallel.
 \item \code{--lexicographic off} forces \tool{T2} to use the original
   \tool{TERMINATOR} method based on disjunctively well-founded transition
   invariants.
 \item \code{--try\_nonterm false} turns off the non-termination prover, useful
   for when such proofs would be unsound due to over-approximating
   pre-processing.
\end{itemize}


\fi

\end{document}